# Dosimetric Study of Lung Modulation and Motion Effects in Carbon ion Therapy for Lung Cancer

M.C. Martire[1,2], L. Volz[1], M. Durante[1,3], C. Graeff *[1,2]

[1] Biophysics, GSI Helmholtzzentrum für Schwerionenforschung, Darmstadt, Germany;

[2] Department of Electrical Engineering and Information Technology, TU Darmstadt, Germany

[3] Institute for Condensed Matter Physics, TU Darmstadt, Germany;

*Author to whom any correspondence should be addressed.

E-mail: xxx@xxx.xx

**Abstract**

Carbon-ion radiotherapy offers superior dose conformity for lung cancer, but its clinical benefit is limited by two physics-driven uncertainties: the interplay between scanned beam and tumor motion, and the modualtion effect of heterogeneous lung tissue that broadens the dose. This work quantifies the individual and combined dosimetric impact of these effects using the GSI in-house TRiP4D treatment-planning system (TPS).

Eighteen lung-cancer patients 4DCT were available from The Cancer Imaging Archive (TCIA). For each patient, a modulation power (Pmod) was assigned to the lung voxel. Three inflated-lung Pmod values were extracted from a Gaussian distribution (200µm±67µm) and an extreme value of 750µm was tested. Interplay doses were generated superimposing the scanned-beam delivery sequence with patient-specific respiratory motion extracted from the 4DCT. Four dose scenarios were computed for each patient: static, static with lung modulation, interplay, and interplay with lung modulation. Target-coverage (D95%, V95%, homogeneity index = HI) and normal-tissue metrics (lung V16Gy, heart V20Gy) were evaluated. The influence of modulation strength, tumour depth, and motion amplitude was examined, and 4D optimization and fractionation were tested as mitigation strategies.

Interplay alone degraded target coverage by 5.2±1.5pp (D95%), 12.1±5.9pp (V95%) and 8.3±2.4pp (HI) (p<0.01). The "Extreme Pmod" (750 µm) produced a minor deterioration of 1.4±0.9pp (D95%), 0.6±1.1pp (V95%) and 0.3±0.9pp (HI). When the extreme modulation was applied to the interplay dose, the plan improved on average by 4.1±1.2pp (D95%), 7.2±4.5pp (V95%) and 5.4±1.5pp (HI) relative to pure interplay. This compensating effect diminished when 4D optimization was employed. Fractionating the treatment averaged out the interplay, leaving the systematic lung modulation as the dominant residual effect.

Interplay is the primary source of dose degradation in scanned carbon-ion lung therapy, whereas lung heterogeneity introduces a smaller, effect that can partially compensate interplay-induced degradations. When motion mitigation strategies reduce interplay, the lung modulation effect becomes prominent.

## 1.    Introduction

Particle Therapy (PT), characterized by highly conformal dose deposition, has developed as a promising alternative to conventional photon therapy [1]. With heavier ions, like carbon ions, the even sharper Bragg Peak (BP) depth dose deposition (DDD) also comes with enhanced radiobiological efficacy. Therefore, carbon radiotherapy (CIRT) has the potential to effectively treat radioresistant tumours, such as locally advanced non-small cell lung cancers (NSCLC) [2]. CRT, because of its physical and biological characteristics, requires extremely precise dose deposition in the target, which is impaired in thoracic treatments by range and energy uncertainties mainly deriving from setup uncertainties, breathing motion and lung tissue inhomogeneities. Lung tissue microstructure and its density variation are responsible for the so-called lung modulation effect, characterized by a substantial broadening of the BP [3], [4]. The superposition of pencil beam scanned motion and tissue motion, cause the interplay effect [5][6] [7], leading to highly heterogeneous dose distributions with sever over- and under-dosage regions in the target.

Intra-fractional motion dose degradation is a well-known problem in the PT community and multiple mitigation techniques (i.e. gating, breath-hold, tracking, rescanning) have been widely investigated and, with the exception on tracking, introduced in clinical practice [8][9],[10] [11]. Patient-specific motion is accounted for in treatment planning with time resolved CT (4DCT), motion included target safety margins [12],[13] and advanced robust optimization (RO) techniques. Integrating motion management at the plan optimization stage, is defined as 4D optimization methods [14], [15], [16].

The modulating effect of heterogenous tissues has been investigated in multiple studies [17], [18], [4], particularly the impact of lung tissue density variations on the deposited dose have been studied in Monte Carlo (MC) simulations and experimentally verified on porous materials or porcine lung samples. Results show these heterogeneities cause DDD degradation in the distal fall-off, possibly leading to target underdosage and normal tissues overdosage in patient treatment plans [3], [19],[20].

The implementation of lung modulation effect in treatment planning is hindered by the standard clinical CT voxel dimension (mm), which does not resolve the lung heterogeneous microstructure [21] but shows the lung tissue as a low-density homogeneous material. Titt et al. [22], Baumann et al. [19] and Ringbaek et al. [20] in their studies developed a model to analytically describe the lung modulation effect on a particle beam. This was implemented in MC based dose calculation. The mathematical model reproduced the DDD modulation by a convolution of the unperturbed DDD with a Gaussian distribution. The kernel of the Gaussian is defined as the Modulation Power (Pmod) quantity in water-equivalent thickness (WET) units.

Paz et al. [23] further improved the model to consider the lung modulation effect in RBE-weighted dose computations and implemented it in the analytical dose calculation algorithm of GSI's in-house TPS TRiP98, which also enabled to consider this effect already in plan optimization. While for a carbon ion beam in homogeneous material, the RBE maximum would coincide well with the BP [24], Paz et al. demonstrated that the RBE maximum shifts to greater depth as a consequence of the modulation caused by tissue heterogeneities. This shift of high-LET particles, further investigated by Zhao et al [25], is potentially harmful for the normal tissues placed at the distal edge of the target.

The described methods were employed in phantom studies [26] and patient CT studies [17],[3], [23] to investigate lung modulation dose uncertainties in patient-representative scenarios. However, the RBE shift and carbon ion RBE-weighted dose distributions degradation have so far only been investigated in the study by Paz et al. and only for a single patient. Additionally, to fully model a realistic clinical scenario, the lung modulation effect needs to be considered also in the context of breathing motion, i.e., considering the interplay effect. To date, a treatment planning study considering both intra-fractional motion and tissue heterogeneities is still missing but is crucial to understand the importance of lung modulation in clinical CIRT of lung tumours. We therefore extended the lung modulation model to integrate it into the 4D dose calculation and plan optimization features of TRiP98 (TRiP4D) [27], [28], allowing us to perform such a dosimetric study.

In this work, we present the isolated as well as the combined effect of both the interplay and lung modulation effect on CIRT treatment plans for a cohort of eighteen patients. 3D and 4D optimizations and dose calculations were performed in TRiP4D investigating the impact of effect-specific parameters such as the strength of the Modulation Power and target depth in lung for the modulation effect, and motion amplitude for the interplay effect, were investigated.

## 2.    Materials and Methods

### 2.1   *Patient Data*

The dataset used in this study was available from "The Cancer Imaging Archive" (TCIA), an open-access database of medical images for cancer research.  Eighteen NSCLC patients 4DCT were selected from the "4D-Lung" collection of TCIA [29]. For treatment planning purposes, only 4DCT images with available contoured volumes of interest (VOI) were selected. 4DCT images were acquired on a 16-slice helical CT scanner (Brilliance Big Bore, Philips Medical Systems, Andover, MA) as respiration-correlated CTs with 10 breathing phases (0% to 90%, phase-based binning) and 3mm slice thickness. A single Radiation Oncologist delineated targets and organs at risk (OARs) in all 4DCTs. Ipsi- and contra-lateral lungs as well as the heart were evaluated for this study. The target position, volume and motion, together with its depth in lung are the main patients characteristics of interest for this study and are listed in **Table 1**. The clinical target volume (CTV) displacement was calculated as the average magnitude of the deformable vector field (DVF) correlating the target voxels from the end-inhale (0% respiratory cycle) to the end-exhale (50% respiratory cycle) phase. In addition, the distance, in the central slice of the reference phase gross tumour volume (GTV), between GTV and the lung contour was calculated in 3DSlicer [30] for each field direction.

| Patient | Target position | Target volume (cc) | Target motion (mm) | Distance in lung (mm) field1/field2 |
|---------|-----------------|--------------------|--------------------|-------------------------------------|
| P1 | RUL | 100 | 2.9 | 59/55 |
| P2 | RUL | 46 | 6.1 | 59/56 |
| P3 | LUL | 196 | 2.4 | 72/65 |
| P4 | LLL | 40 | 6.0 | 64/48 |
| P5 | RUL | 167 | 3.7 | 37/37 |
| P6 | LUL | 47 | 3.7 | 0/0 |
| P7 | RUL | 19 | 5.5 | 94/61 |

| P8 | RLL | 23 | 8.3 | 44/46 |
|---|---|---|---|---|
| P9 | RUL | 74 | 3.1 | 32/90 |
| P10 | LUL | 23 | 3.4 | 10/65 |
| P11 | RUL | 15 | 1.1 | 40/42 |
| P12 | Mediastinum | 393 | 4.1 | 15/83 |
| P13 | RML | 75 | 7.5 | 23/81 |
| P14 | RLL | 118 | 11.9 | 0/82 |
| P15 | RUL | 32 | 1.1 | 0/0 |
| P16 | Mediastinum | 85 | 5.6 | 0/0 |
| P17 | RLL | 218 | 9.6 | 16/96 |
| P18 | RUL | 8 | 3.9 | 6/65 |

Table 1. Patients characteristics. Target position in lung, volume and motion amplitude, as well as target distances in lung for filed 1 and filed 2 directions are listed for each patient. The target motion amplitude was obtained from deformable vector fields (DVF) amplitude between end-exhale and end-inhale breathing phases. The depth in lung was computed from the central slice of the GTV in the reference phase CT. RUL=right upper lobe, RLL=right lower lobe, RML=right middle lobe, LUL=left upper lobe, LLL=left lower lobe.

## 2.2 Treatment Planning

Carbon ion intensity modulated particle therapy (IMPT) plan optimization and dose calculations were performed with the GSI in-house TPS TRiP4D [31], [32],[27], [28]. To estimate the carbon ion relative biological effectiveness (RBE) we used LEM I and a global alpha/betha ratio of 2. 3D-optimized plans were computed on the reference phase CT (0% respiratory cycle) for a dose of 3GyRBE/fraction to the geometrical ITV after density override. A full 4D optimization strategy on the CTV [33] was also employed for eight patients with target motion exceeding 5mm [34],[33], [10], [14]. For both optimization strategies, two-field coplanar plans were considered. All plans were robustly optimized with 3.5% range uncertainty and setup uncertainty varying from 3 to 7 mm, to obtain patient-specific margins. Larger margins were used to achieve proper target coverage in presence of larger target motion or volume.

In addition to static 3D doses, calculated on the reference phase CT, TRiP4D allows for 4D and dynamic 4D (D4D) dose [15] computation to account for target motion amplitude alone and the interplay effect, respectively. In addition, the implementation of lung modulation degradation model in TRiP4D [23], allows for RBE-weighted dose calculation reproducing the plan degradation caused by microscopic lung heterogeneities.

### 2.2.1 4D dose calculations

RBE-weighted D4D doses simulate the superposition of scanned beam motion and patient breathing motion, resulting in the interplay effect doses. Plan delivery is simulated using an in-house beam delivery simulator Richter 2013], reproducing the accelerator and beamline characteristics of the Heidelberg Ion Beam Therapy Center, Marburg Ion Beam Therapy Center, and Shanghai Proton and Heavy Ion Therapy Center. Since the original patient breathing motion was not available, a motion function, with 10s breathing period, was used as surrogate while the motion amplitude was directly extracted from patient specific 4DCT. This anatomical information is obtained computing deformable image registration (DIR) between the reference CT and all other phases. The DIR is computed in Plastimatch [35][35] using three bspline registration stages, without preliminary rigid registration.

4D doses, instead, were evenly distributed over the 10 phases of the 4DCT and then accumulated on the reference CT through dose deformation. The beam delivery structure is not included, and the result reflects what would be obtained with an infinite number of rescan.

### 2.2.2  Lung modulation doses in TRiP4D

Heavy ions traversing lung tissue heterogeneity undergo energy and range straggling because of the multiple paths particles can take in the locally changing density of a material. This leads to modulation of the reference BP, resulting in lower dose deposition inside the target and a broadened dose distribution in tissue located at the distal edge of the target [36].

The lung tissue dose degradation and its effect can be mathematically described by a convolution between the unperturbed dose distribution $b_0(z)$, traversing a binary voxelised geometry, and a gaussian distribution P(t'|σ,t) describing the probability that a certain path had a (WET) $t'$.

$$P(t' \mid \sigma, t) = \frac{1}{\sqrt{2\pi\sigma^2}} exp\left(-\frac{(t'-t)^2}{2\sigma^2}\right) \qquad 1]$$

Where t is the average WET of the voxelised geometry over the whole CT in beams eye view and σ is related to the heterogeneity of the specific material.

Therefore, to obtain the modulated dose distribution $b_*(z)$ at each depth z, the following convolution is computed

$$b*(z) = (P(t' \mid \sigma, t) * b0)(z) = \int_{-\infty}^{\infty} P(t' \mid \sigma, t) \, b_0(z+t') \, dt' \qquad 2]$$

This mathematical model results in the definition of the Modulation Power (Pmod) as new physical quantity to describe the material specific modulation in units of WET.

$$Pmod = \frac{\sigma^2}{t} [\mu m] \qquad 3]$$

Pmod depends on the size of the microscopic structure, leading to stronger degradation for increased structure size [19].

The implementation of this model into TRiP98 is described in detail in the work by Paz et al. [23]. Briefly, physical modulated doses are computed applying the convolution for each depth z. Then, the RBE-weighted doses are obtained convolving the dose-averaged α, β, and LET distributions of the reference BP with the same Gaussian function. In the TPS, a global constant Pmod value is assigned to each CT voxel identified as lung parenchyma (-900<HU<-500), and this allows to construct the Gaussian modulation function, from the inverse formula of equation 3.

Therefore, modulated dose calculations were computed using two Pmod strategies:

1. Pmod of inflated lung, experimentally validated in previous studies [37], [4], [3], [38] and described by a normal distribution (200 μm ±67 μm).
   To have a more realistic assessment of the lung modulation impact, for each patient, three dose calculations were computed with a constant Pmod value sampled from this normal distribution. We refer to these values as "Gaussian Pmod".

2. An "Extreme Pmod" value of 750 µm was selected to show an upper limit of a possible effect

Interplay and lung modulation effect were investigated bot in isolation and combined, to understand the individual contributions. Motion related dose degradations were quantified by comparing the 3D reference plan and the D4D doses, while the comparison between 3D reference plans and 3D modulated doses enables to evaluate for the lung modulation effect. The latter comparison was performed both for "Gaussian Pmod" and "Extreme Pmod" case.

To quantify the combined effect, the lung modulation model implementation was extended to the 4D features of TRiP4D. Briefly, the convolution of dose-averaged $\alpha$, $\beta$, and LET distributions of the reference BP with the Gaussian function is computed in each voxel for the specific depth z and the assigned breathing motion. Therefore, it was possible to compute D4D RBE-weighted doses with the superposition of the lung modulation effect. Comparing these doses with the unmodulated D4D plans allowed to assess the combination of both effects, mimicking what would happen in a real clinical scenario.

### 2.3 *Data Analysis*

Plan evaluation was performed both qualitatively and quantitatively. Volumetric dose distributions displayed in 3DSlicer were used for qualitative evaluation, while dose volume histograms (DVH) and extracted dosimetric metrics were employed for quantitative analysis. Metrics of interest are D95% and the mean dose D50% (prescribed dose percentage delivered to 95% and 50% of the target volume), V95% (target volume percentage receiving at least 95% of the prescribed dose) and homogeneity index (HI= D5% - D95%) to quantify the target coverage, and V20Gy and V16Gy were used as the reference metrics for the heart and the lung, respectively. The organs at risk (OARs) metrics derive from lung V20Gy and heart V25Gy used in the clinic for the standard 2GyRBE/fraction in 30 fractions treatment and converted to the corresponding values for 3GyRBE/fraction in 20 fractions used in this study. The plan was considered clinically acceptable for D95% and V95% greater than 95%, and an HI lower than 5%, as well as heart V20Gy lower than 10%, and lung V16Gy less than 30% [39]. Target DVH curves and dosimetric metrics were calculated on the CTV, namely the reference state GTV expanded by the isotropic patient-specific margins.

Target coverage differences were assessed with paired T-tests and Wilcoxon signed rank tests, considering p<0.05 as statistically significant. In particular, when comparing 3D and 4D optimization strategies, because of the reduced sample size, the Wilcoxon signed rank test was used.

A linear correlation analysis was performed to investigate the dependency of lung modulation degradation and target depth in lung. The target coverage (i.e. D95%) difference between reference 3D and modulated 3D doses were plotted as a function of target depth in lung. For each patient, the average target depth in lung of the two fields (average of the values in Table 1), was used. Linear correlations were calculated for the "Gaussian Pmod" and the "Extreme Pmod" cases.

To investigate the interplay and lung modulation combined effect, the target coverage degradation was investigated as a function of motion amplitude, both for unmodulated (Pmod=0µm) and modulated (Gaussian and Extreme Pmod) interplay doses. Therefore, a linear correlation was computed between the difference of reference 3D to D4D doses D95% and the target displacement.

The superposition of interplay and modulation effects was further investigated in terms of target coverage convergence over multiple fractions. For the patient with the highest motion amplitude, P14, interplay doses were calculated selecting each of the 4DCT phases as the starting phase of the motion surrogate, for a total of 10 different D4D doses. These doses were summed up to reproduce 20 fractions, for a total cumulative dose of 60GyRBE. Four treatment courses were computed randomly sampling the D4D doses from the available data, and for each fraction, the averaged values are displayed. The above-described method was performed both for unmodulated (Pmod=0µm) and modulated ("Extreme Pmod") interplay doses, to evaluate the difference in plan convergence. The convergence speed is defined as the number of fractions necessary to achieve the metric's plateau (+/-1%), which ideally is above/below the minimum/maximum clinically acceptable limit for the specific metric (i.e. 95% for D95).

## 3. Results

### 3.1 Isolated effect study

#### 3.1.1 Interplay effect

After patient-specific 3D plan optimization, 3D, 4D and D4D doses were calculated as described in Section 2.2, and plan quality evaluation was carried out as described in Section 2.3. For each patient, target and OARs metrics of 3D and D4D doses were compared. P14 and P15, characterized by the maximum (11.9 mm) and minimum (1.1 mm) motion amplitude, were chosen as representative cases, and shown in **Figure 1**. Patient P14 showed more severe target dose degradation, quantified by D95% and V95% decrease of 5.6 pp and 13.9 pp, respectively, and HI increase by 8.8 pp. Instead, patient P15 D95% and V95% diminished by 2.7 pp and 3.1pp, respectively, and HI increased by 4.9 pp.

On average, the interplay target dose degradation, calculated for the whole patient cohort, is quantified by a decrease of 5.2±1.5 pp and 12.1±5.9 pp for D95% and V95%, respectively, and 8.3 ±2.4 pp increase for HI. These differences resulted statistically significant. Interplay dose deteriorations mainly affect the target, without significant changes to OARs doses. Indeed, on average the lung V16Gy increased by 1.1±2.6 pp and the heart V20Gy decreased by 0.3±0.3 pp.

When the target displacement is considered alone and 4D doses are compared to reference 3D ones, D95%, V95%, and HI worsened of 1.7±1.2 pp, 1.4±2.1 pp and 0.6±1 pp, respectively. Notably, the plan deterioration magnitude is much larger when the interplay effect is considered.

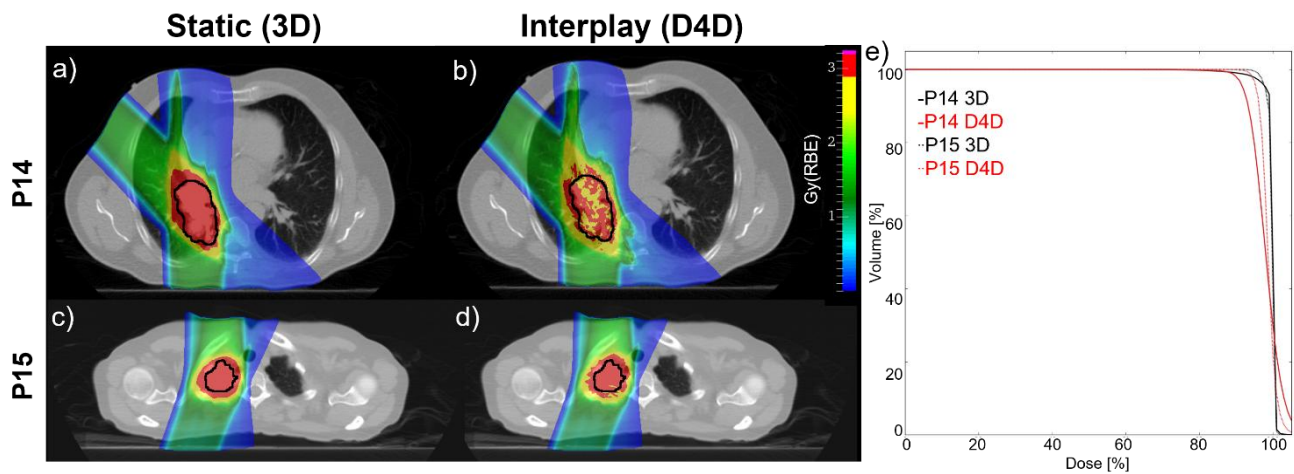

Figure 1. 3D (a,c) and D4D (b,d) plans for patient P14 (a,b) and P15 (c,d), characterized by the maximum and minimum target motion amplitude. DVH (e) of static (black) and interplay (red) cases are also compared for patient P14 (solid line) and P15 (dotted line). The interplay effect magnitude is larger for higher target motion but severely impacts the plan quality also for small motion. D4D= Dynamic 4D

#### 3.1.2 Lung modulation effect

Starting from the same 3D-optimized plans, modulated dose calculations were computed as described in Section 1.2.2 and compared to the reference 3D unmodulated doses, for each patient. In **Figure 2**, the DVH of patient P4 is shown as representative case. The OARs (ipsi-/contra-lateral lung and heart) curves show no differences, while increased variations are visible when comparing CTV lines of the reference and modulated cases. Patient P4 "Extreme Pmod" dose degradation is quantified by 2pp, 1.2pp decrease of D95% and V95% respectively. No changes are present for the HI metric. For the "Gaussian Pmod" doses, instead, patient P4 shows a decrease of 1.6pp and 0.7pp for D95% and V95%, respectively. The modulation effect, smoothing the dose distribution and reducing hot spots, leads to steeper gradient in the DVH around a reduced D50% value. Patient P4 D50% decreases of 2pp and 1.7pp, for the "Extreme Pmod" and "Gaussian Pmod", respectively.

The "Extreme Pmod" target degradation, calculated for the whole patient cohort, is quantified by an average decrease of 1.4±0.9pp and 0.6±1.1pp for D95% and V95%, respectively, and 0.3±0.9pp increase for HI. The mean dose instead decreased on average by 1.1±0.7pp. D95% and D50% differences resulted statistically significant. The depth dose profile degradation caused by lung inhomogeneities, not only diminished the dose inside the target, but also shifts it in depth. However, this only affected normal tissues distal to the tumour. In fact, when the "Extreme Pmod" doses were compared to the reference ones, no meaningful V16Gy and heart V20Gy changes were found.

Together with the assigned modulation strength, the lung modulation effect highly depends on the distance particles traverse in lung tissue. To quantify this dependency, a linear correlation analysis was computed as described in Section 1.3 and shown in **Figure 3**. For each patient, the reference and modulated target coverage difference is evaluated through D95% and plotted against the average target depth in lung for the two fields directions. The analysis resulted in a moderate positive correlation quantified by a correlation coefficient of 0.4 ($p<0.05$) and 0.5 ($p<0.05$), for "Gaussian Pmod" and "Extreme Pmod" doses, respectively.

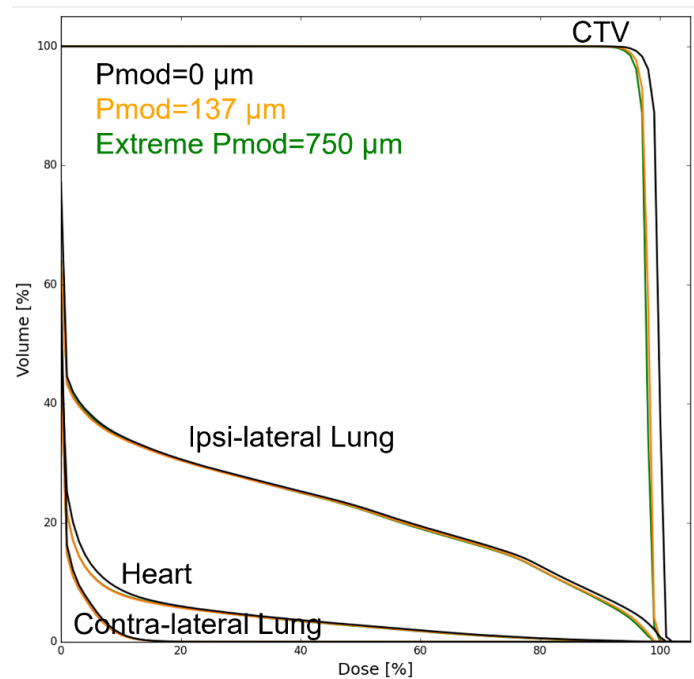

Figure 2. Patient P4 static plan DVHs for reference (black), Gaussian Pmod (orange) and Extreme Pmod (green) cases. OARs (Heart, ipsi-/contra-lateral lung) and target (CTV) lines are plotted. Lung modulation minorly affects target coverage, while diminishing the mean target dose. Small differences are visible for Pmod values of 137 µm and 750µm. OARs show no differences when unmodulated and modulated plans are compared. CTV=Clinical Target Volume, Pmod=Modulation Power

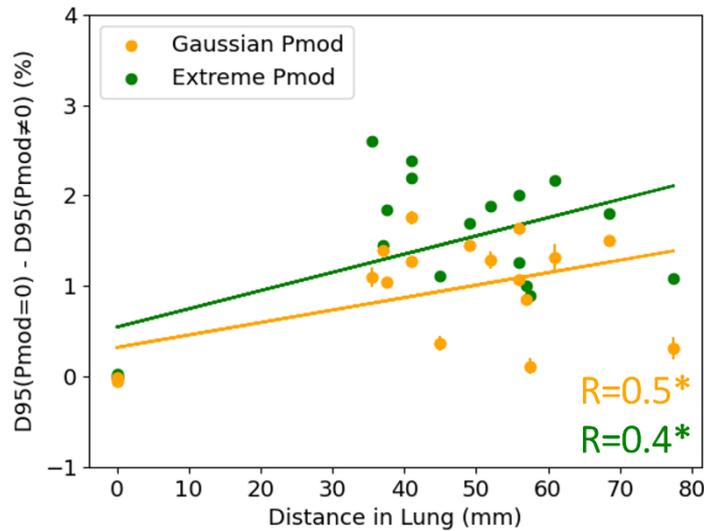

Figure 3. Linear correlation between D95% of unmodulated (Pmod=0) and modulated (Pmod≠0) 3D plans difference and target depth in lung. The correlation is computed for the Gaussian Pmod (orange) and Extreme Pmod (green) case. Gaussian Pmod values are plotted as average (dot) and standard deviation (bar) over three values. Correlation Coefficient (R) are written, and statistical significance is marked with a star (*).

### 3.2    *Combined effect study*

To reproduce a realistic clinical scenario, the superposition of interplay and lung modulation effects was investigated. In **Figure 4**, reference 3D, unmodulated D4D and modulated D4D dose distributions are displayed in axial view for patient P14. The dose broadening caused by tissue inhomogeneities smooths the dose reducing hot and cold spots inside the target. This is qualitatively shown comparing Figure 4.b and 4.c and confirmed by the quantitative analysis. In fact, the modulated ("Extreme Pmod") interplay dose degradation, calculated on the whole patient cohort, is quantified by an average increase of 4.1±1.2pp and 7.2±4.5pp for D95% and V95%, respectively, and 5.4±1.5pp decrease for HI, compared to the D4D dose without modulation. The "Gaussian Pmod" modulated interplay doses, showed an average increase of 0.8±1.9pp and 3.7±7.5pp for D95% and V95%, and 1.9±2.9pp decrease for HI, compared to unmodulated D4D doses. Both for Extreme and Gaussian Pmod, the differences in D95% and HI statistically significantly improved compared to the D4D doses without modulation, but also statistically significantly better than the doses for each of the effects in isolation. No meaningful lung V16Gy and heart V20Gy changes were found.

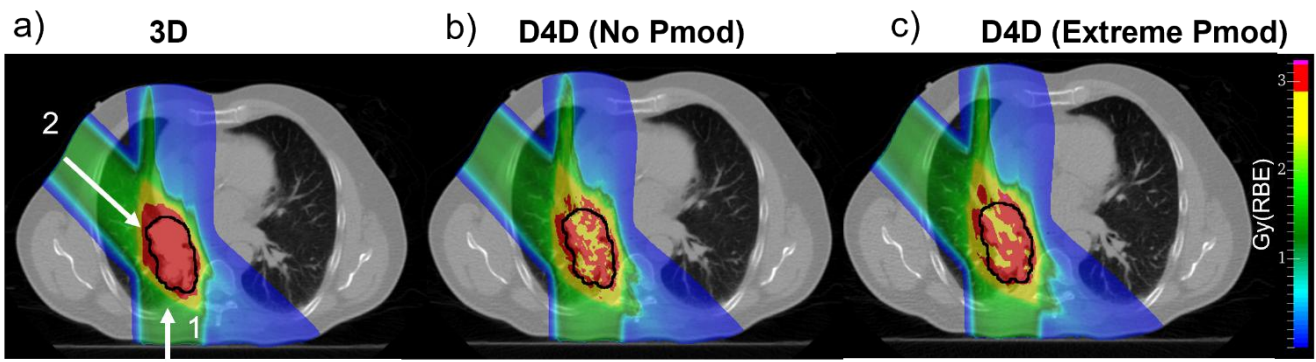

Figure 4. Patient P14 3D (a), unmodulated D4D (b) and Extreme Pmod D4D (c) dose distributions are compared. The combination of lung modulation and interplay effect partially recovers interplay dose distortions. D4D= Dynamic 4D, Pmod=Modulation Power.

The target motion amplitude is the main factor contributing to the interplay effect magnitude. Therefore, we computed a correlation analysis between the target coverage degradation and the target displacement, as described in Section 1.3. The linear correlation result is shown in **Figure 5**. A correlation coefficient of 0.7 (p<0.05) and 0.6 (p<0.05) describes the positive correlation between motion amplitude and unmodulated and modulated D4D doses, respectively. When the "Extreme Pmod" is applied to interplay doses, the correlation coefficient decreases to 0.4 (p> 0.05). Therefore, the target displacement has a diminished influence on dose degradation when also considering of the lung tissue modulation effect, and the influence further decreases for larger the modulation values.

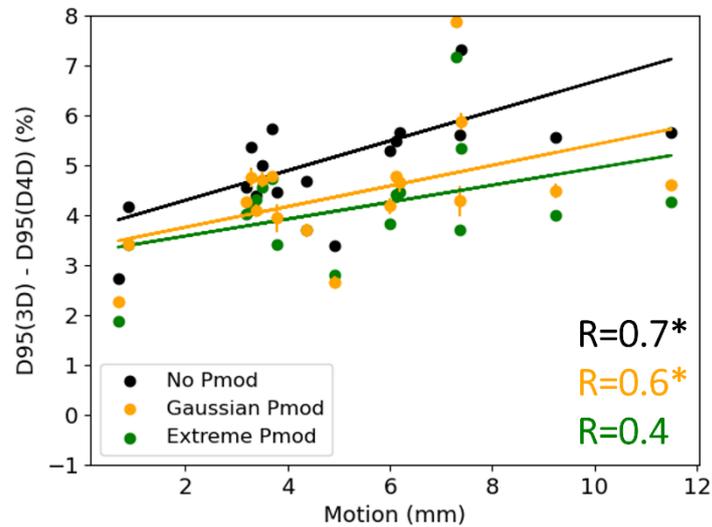

Figure 5. Linear correlation between D95% of 3D (black) and D4D (orange/green) difference and target motion amplitude. The analysis is computed for the Gaussian Pmod (orange) and Extreme Pmod (green) case. Gaussian Pmod values are plotted as average (dot) and standard deviation (bar) over three values. Correlation Coefficient (R) are written, and statistical significance is marked with a star (*).

In **Figure 6** the target coverage convergence over 20 fractions is displayed in terms of D95%, V95% and HI metrics, both for unmodulated and modulated ("Extreme Pmod") interplay doses. The lung modulation contribution, reducing hot and cold spots inside the target, positively affects the target coverage at the beginning of the treatment. However, already at the third fraction, unmodulated and modulated D4D plans reach comparable quality.

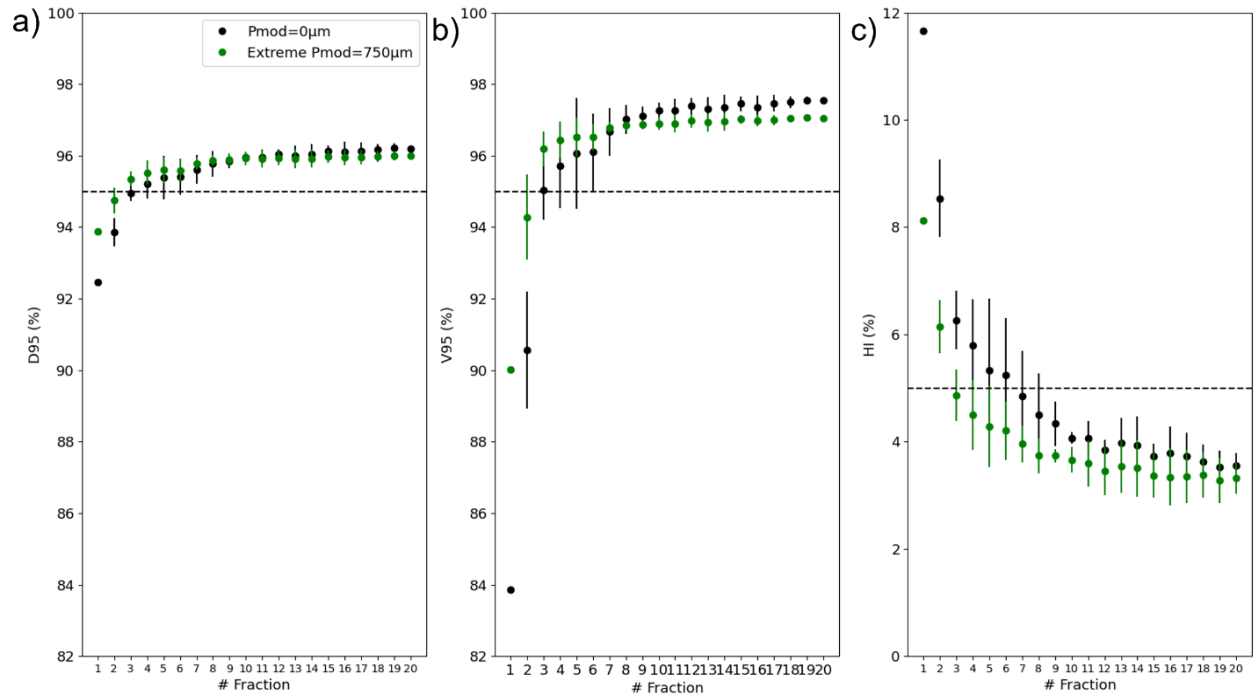

Figure 6. Patient P14 target D4D dose convergence over 20 fractions for unmodulated (black) and Extreme Pmod (green) plans. D95% (a), V95% (b) and HI (c) are plotted. Four fractionation treatments are simulated and for each fraction the average (dot) and standard deviation (bar) values are plotted. The minimum clinically acceptable values of 95% and 5% are marked with dashed lines for D95% and HI, respectively.

### 3.3    *4D optimization*

As described in Section 2.2, for patients with motion amplitude greater than 5mm, 4D-optimized plans and D4D doses were computed with and without considering the lung modulation effect. In **Figure 7** interplay doses computed from 3D and 4D-optimized plans are shown for patient P14.

The sub-group of eight patients exhibits on average an increase of 0.3±2.6 pp, 3.7±8.6 pp and 1.2±0.8 pp for D95%, V95% and D50%, respectively, and an increase of 2±3.2 pp for HI. When comparing the "Extreme Pmod" D4D doses of the 3D and 4D optimized plans, an average increase of 0.3±2.6 pp, 2.1±7.9 pp and 1.2±0.8pp occurs for D95%, V95% and D50%, respectively, while HI shows an increase of 1.1±2.8pp.

In addition, when 3D and 4D optimization strategies are compared in terms of 4D doses, an average increase of 0.7±2.8, 0.5±3.7pp, 0.5±1.3 pp and 0.4±2.6pp occurred for D95%, V95%, D50% and HI.

None of these differences resulted statistically significant, and a large inter-patient variability was found when comparing 3D and 4D optimization strategies. However, the effect of lung modulation reducing interplay degradations is still visible for 4D optimized plans. Quantitatively, a statistically significant increase of 1.6±2.9 pp, 4.8±7.6 pp occurs for D95% and V95%, respectively, while HI shows a decrease of 4.6±3.2pp.

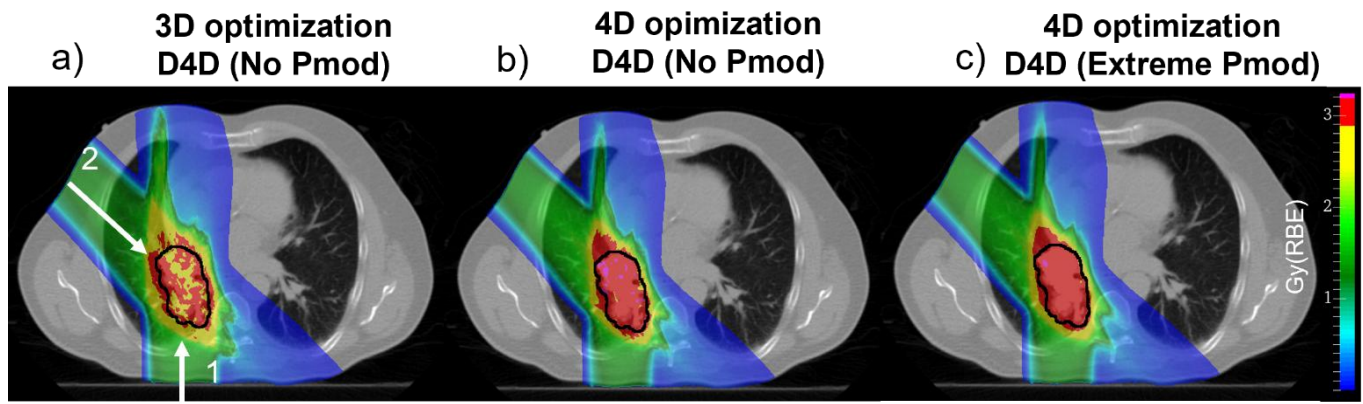

Figure 7. Patient P14 D4D dose distributions compared after 3D (a) and 4D optimization (b,c). The impact of Extreme Pmod value on the D4D plan is shown (c). D4D= Dynamic 4D, Pmod=Modulation Power.

### 3.4    *Results Overview*

**Figure 8** shows the results overview. In each row the following studies, previously described, are displayed in terms of the target metrics D95%, D50% and HI:

- Interplay effect (8.a): 3D doses (reference) compared to D4D doses, without including the lung modulation effect (Pmod=0μm)

- Lung modulation effect (8.b): 3D dose without Pmod (reference) compared to "Gaussian" and "Extreme Pmod" static doses

- Interplay and lung modulation combined effect (8.c): D4D doses without Pmod (reference) compared to "Gaussian" and "Extreme Pmod" interplay doses

- Interplay effect in 4D optimized plans (8.d): D4D doses after 3D optimization (reference) compared to 4D optimization D4D doses

- Interplay and lung modulation combined effect in 4D optimized plans (8.e): D4D doses without Pmod (reference) compared to the "Extreme Pmod" D4D doses, after 4D optimization.

For each comparison, patient-specific metrics are plotted together with the difference distributions. The reference case is shown in blue, and the comparison case is shown in orange and/or green.

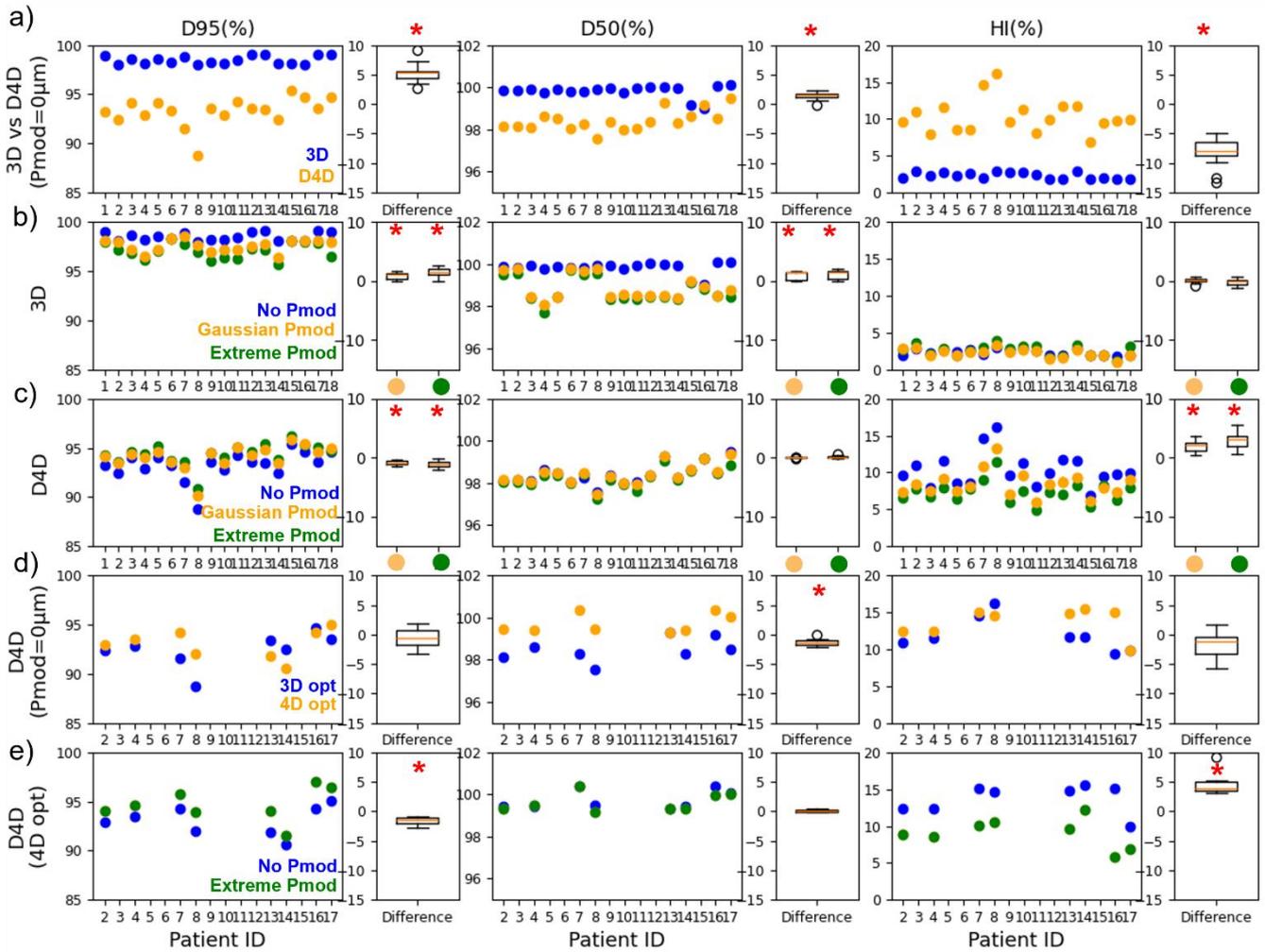

Figure 8. Results overview. Patient-specific D95%, D50% and HI metrics compared for 3D and D4D doses (a); reference Gaussian and Extreme Pmod 3D doses (b); reference Gaussian and Extreme Pmod D4D doses (c); D4D doses from 3D and 4D optimized plans (d); unmodulated and Extreme Pmod D4D doses of 4D optimized plans (e). Metric values differences between the reference case (blue) and the comparison case (orange and/or green) are plotted with boxplots, and statistical significance (p<0.05) is marked with a red start (*). D4D=Dynamic 4D, Pmod=Modulation Power, opt=optimization.

## 4.    Discussion

In this study, we evaluated the clinical impact of interplay and lung modulation, both as isolated effects and in combination, using a cohort of eighteen patients available from TCIA. 3D and 4D dose optimizations and calculations were carried out with TRiP4D. Effect-specific parameters were analysed, such as the Pmod value and target depth in lung for the modulation effect, amd motion amplitude for the interplay effect. Additionally, by comparing unmodulated and modulated interplay doses, we could show that lung modulation indeed has a dampening effect on interplay, partially restoring the deteriorated dose. When considering other effects mitigating interplay, such as 4D-optimization or fractionation, the mitigating effect of modulation becomes smaller and eventually persists as a small negative effect.

### 4.1    Intra-fractional motion in lung cancer treatments

In CRT of thoracic patients, multiple factors produce range uncertainties, such as setup errors, and tissue displacement deriving from breathing motion. The latter strongly affects the finite range of ions, especially when superimposed to the actively scanned beam motion [5]. The result is severe dose distortion mainly affecting the target with hot and cold spots distributed inside the volume. Jakobi et al. [34] investigated the interplay effect of forty free-breathing lung cancer patients, showing that an average absolute V95% reduction of 2% and 13% occurs for patients with motion amplitude smaller and bigger than 5mm, respectively. This is in accordance with what described in Section 3.1.1, where P14 and P15 are reported as maximum and minimum target displacement cases. The direct correlation between interplay effect and motion amplitude is shown in our work and was demonstrated in multiple studies [34], [40]. Thus, motion amplitude is used as main variable in clinical practice to decide in favour or against the application of motion mitigation strategies [8]. In particular, if the amplitude exceeds the 5mm motion management techniques are considered clinically necessary [41]. Therefore, we used the same criterion to create the sub-group of eight patients for which the 4D optimization technique was computed. Figure 8.d compares 3D and 4D optimized interplay doses, showing that a 4D optimizations can mitigate interplay effect degradation. However, inter-patient variability remains large.

The random nature of motion-induced dose degradation restores plan quality after multiple fractions [34],[42], [43]. Because of its dependency on the variable breathing starting phase, motion period and amplitude, the interplay effect is averaged out over multiple fractions of a treatment course. This is shown in **Figure 6** for patient P14.

## 4.2    Lung modulation in carbon ion therapy

The effect of lung tissue heterogeneities on particles range and energy has been widely characterized through MC simulations as well as phantom and porcine lung tissue experiments. The dose deposition changes, mainly causing target underdosage and dose broadening at the distal fall-off, potentially impact the plan quality of lung patients treated with CRT. However, range errors deriving from setup and motion uncertainties have been described as major source of dose degradation in particle therapy for thoracic tumours [44]. In contrast, as shown in Section 3.1.2, the impact of lung modulation effect is responsible for minor changes in target coverage and dose to the OARs. Baumann et al. [4] retrospectively investigated proton plan degradation of five lung cancer patients, reporting a D95% decrease of maximum 5% when the extreme Pmod value of 800μm is applied in the MC dose calculations. In addition, they investigated the maximum range uncertainty of modulated cases when single field, or multiple fields plan are optimized. With an increased number of fields, the range uncertainty decreases. Winter et al. in a retrospective proton therapy study [3] investigated the lung modulation effect in ten patients, applying Pmod values of 256 μm and 750 μm as realistic and extreme case, respectively. They found a D95% worsening in the order of 0.8% for centrally located target and 1.6% for deep seated tumour. This is in accordance with the average D95% decrease of 1.4±0.9pp we found when applying "Extreme Pmod" value to 3D doses and with the positive correlation between modulation magnitude and tumour depth in the lung (see Figure 3). Figure 8.b shows that for patient P5, P15 and P16, characterized by both fields traversing no lung tissues, the modulated and unmodulated plan metrics are exactly the same, independent on the Pmod value. In addition, Winter et al., observed a D50% decrease as demonstration of target underdosage.

Paz et al. showed treatment planning case study for which the lung modulation effect was included in RBE-weighted dose calculation. The plan was designed to maximise the particles path in lung and resulted in a D95% decrease of 1.5% and 3.2% for Pmod values of 256μm and 750μm, respectively. The dependency of lung modulation dose degradation and tumour depth in lung, and its effect on the mean dose D50% was also confirmed in the CT-based lung tumour phantom study of Flatten et al. [26]. Notably, in a real clinical scenario, the typical plans would strive to minimize the particles path in lung tissue simply to reduce lung dose. It should also be noted that large airway structures, responsible for increased dose degradation, are resolved in the clinical CT and accounted for in the treatment plans.

## 4.3    Combined Motion and lung modulation effects

The higher impact of intra-fractional uncertainties on the final dose distribution, with respect to lung modulation, have been demonstrated in this study in accordance with previous works. Nevertheless, for a comprehensive analysis of a real clinical scenario, the combination of both effects has to be considered. For this purpose, we investigated for the first time the superposition of both effects and its impact on the RBE-weighted dose as it would occur in the patient.

Our results demonstrate that the dose smearing and broadening caused by the tissue inhomogeneities, so far considered as a purely negative effect, partially compensates for target underdoses and overdoses caused by the interplay effect. This is shown through D95% and V95% improvement of about 1pp and 5pp on average in our patient cohort. While still, for twelve of the eighteen patients, D95% and V95% remained below the minimum acceptable threshold of 95% without any other motion compensation, the combination of interplay and modulation effect was always a net positive. We also demonstrated the diminished correlation between target coverage (D95%) and motion amplitude when the lung modulation is applied to the interplay doses. In particular, for the "Extreme Pmod" case the correlation coefficient decreases below 0.5 and it is no longer statistically significant (see Figure 5). While this result may seem surprising at first, it is not entirely unexpected. The interplay effect is greater for sharper beams than broader ones, due to the higher localized dose deposit.

We want to highlight that modulation strength of heterogenous and porous materials have been proposed by Ringbaek et al. [20] as a ripple filter substitute in particle therapy. Especially for carbon ions, the extremely sharp BP needs to be broadened before it is delivered to the target to achieve a more homogenous dose. Our results confirm that the porous structure of lung tissue acts as an additional dose modulator on carbon ion DDD. It is likely that this compensating effect is less relevant for protons, as their BP is less sharp especially for deep-seated targets.

Finally, to embed these results in the context of a realistic course, we simulated the cumulative dose over twenty fractions for the reference D4D and the modulated doses. The result shows that the above-mentioned plan improvement caused by the superposition of the two effects is prominent at the beginning of the treatment but almost fades already at the third fraction. In particular, for the V95% metric, the unmodulated plan at fraction 20 is superior to the modulated plan.

This demonstrates that with the averaging of interplay over multiple fractions, the motion degradation diminishes while the lung modulation effect degradations, with its systematic nature, persist. Indeed, the effect of tissue heterogeneities on target underdosage occurs systematically at every fraction and is not compensated over multiple deliveries as are randomized range uncertainties. This can be observed in Figure 6, where fractions restore the motion-induced loss of coverage, but both D95 and V95 show a small but persistent reduction for the Extreme Pmod calculation. It would be interesting to analyse this also in the context of anatomical variations, and fraction-wise varying modulation power. However, the fluctuation of Pmod over multiple fractions has not yet been reported, and hence, this was beyond the scope of the presented manuscript.

Overall, our findings suggest that with a better mitigation of the interplay effect, for example with advanced mitigation strategies, lung modulation degradations become more relevant and require additional attention. Among the motion mitigation methods, 4D optimization strategies, including breathing motion scenarios at the planning stage, have been proposed [14].

## 4.4    Study limitations

This study has some limitations. Firstly, to reproduce the modulation effect in dose calculations, a constant Pmod value is assigned to each CT voxel. This only approximates the lung tissue modulating strength that varies with different diameter of airway components. A patient-specific modulation map could assign to each lung structure a different Pmod value. For example, a method has been proposed by Flatten at al [45] to extract patient specific Pmod directly from the HU histogram of clinical CT. In our work, we mitigated this limitation with the use of a Gaussian Pmod concept. For each patient three modulated doses were computed, randomly extracting the Pmod value from a normal distribution. The average result over these repetitions was then used for the analysis.

Secondly, the same motion patter surrogate was used for interplay dose calculations, while ideally a patient-specific motion pattern could be recorded. However, to properly simulate irregular motion, continuous image series, such as synthetic CTs from 4DMRI, or at least repeated 4DCTs would be required, but were not available for the cohort [46].

Ultimately, the study is limited by the moderate number of patients, but their characteristics are distributed broadly enough to account for multiple variables and derive conclusive results.

## 5.        Conclusions

   In this study, we investigated CRT in eighteen lung cancer patients and analysed the main factors contributing to plan degradation. We separately assessed the effects of interplay and tissue heterogeneity modulation using D4D and 3D dose calculations, respectively, and found that motion-induced uncertainties have a much greater impact on plan quality. The modulation effect of lung heterogeneities plays only a minor role in overall plan degradation, but since it occurs consistently at every fraction, its relative importance increases with the improved motion control offered by modern mitigation strategies.

For the first time, we also examined the combined influence of both effects to reproduce a realistic clinical scenario. We observed that carbon ion BP modulation caused by lung tissue heterogeneities spreads the dose within the target, partially compensating for the under- and over-dosage produced by the interplay effect. However, due to its stochastic nature, interplay degradation becomes negligible after multiple fractions, meaning that the combined effect ultimately cancels out.